\documentclass[aps,pre,twocolumn,superscriptaddress,showpacs]{revtex4}%

\usepackage{amsmath}
\usepackage{amssymb}
\usepackage[dvips]{graphicx}
\usepackage{colordvi}
\usepackage{latexsym}
\usepackage{amsfonts}
\usepackage{times}

\newcommand{\be}{\begin{equation}}
\newcommand{\ee}{\end{equation}}
\newcommand{\bea}{\begin{eqnarray}}
\newcommand{\eea}{\end{eqnarray}}
\newcommand{\beas}[1]{\begin{subequations}\label{#1}\bea}
\newcommand{\eeas}{\eea\end{subequations}}
\newcommand{\width}{8cm}

\newcommand{\MR}{\Ree}
\newcommand{\MI}{\Imm}

\renewcommand{\k}{\mbf{k}}
\newcommand{\mbf}[1]{\mbox{\boldmath{$#1$}}}
\newcommand{\bfsigma}{\mbf{\sigma}}
\renewcommand{\u}{\mbf{u}}
\newcommand{\q}{\mbf{q}}
\newcommand{\e}{\mbf{e}_k}

\newcommand{\ave}[1]{\left\langle#1 \right\rangle}
\renewcommand{\r}{\mbf{r}}
\newcommand{\Ree}{{{\cal R}}}
\newcommand{\Imm}{{{\cal I}}}
\newcommand{\Mk}{{\cal M}_k}
\newcommand{\Tk}{{\cal T}_k}
\newcommand{\Tmk}{{\cal T}_{-k}}

\newcommand{\stl}[1]{{\hspace{0.2em}
\stackrel{ \setbox0=\hbox{\hspace{0.06em}$\displaystyle
#1$\hspace{0.06em}} \setbox1=\hbox{\vrule width\wd0 height0.08ex
depth0pt} \vrule width0.08ex height0.08ex depth0.475ex \box1 \vrule
width0.08ex height0.08ex depth0.475ex } {#1}\hspace{0.2em}}}

\begin{document}

\title{Hyperbolicity of exact hydrodynamics for three-dimensional linearized Grad's
equations}

\author{Matteo Colangeli}\email{matteo.colangeli@mat.ethz.ch}
\affiliation{ETH Z\"urich, Department of Materials, Polymer Physics,
CH-8093 Z\"urich, Switzerland}

\author{Iliya V. Karlin \footnote{Corresponding author}}\email{karlin@lav.mavt.ethz.ch}
\affiliation{ETH Z\"urich, Aerothermochemistry and Combustion
Systems Lab, CH-8092 Z\"urich, Switzerland}

\author{Martin Kr\"oger}\homepage{www.complexfluids.ethz.ch}
\affiliation{ETH Z\"urich, Department of Materials, Polymer Physics,
CH-8093 Z\"urich, Switzerland} \affiliation{Materials Research
Center, ETH Z\"urich, Wolfgang-Pauli-Str.~10, CH-8093 Z\"urich,
Switzerland}


\date{\today}

\begin{abstract}

We extend a recent proof of hyperbolicity of the exact (to all
orders in Knudsen number) linear hydrodynamic equations [M.
Colangeli et al, Phys. Rev. E (2007)] to the three-dimensional
Grad's moment system. A  proof of an $H$-theorem is also
presented.

\end{abstract}

\maketitle

\section{Introduction}

Derivation of hydrodynamics from a microscopic description is the
classical problem of physical kinetics. The Chapman-Enskog method
\cite{chapman} derives the solution from the Boltzmann equation in a
form of a series in powers of Knudsen number $\varepsilon$, where
$\varepsilon$ is a ratio between the mean free path of a particle
and the scale of variations of hydrodynamic fields. The
Chapman-Enskog solution leads to a formal expansion of stress tensor
and of heat flux vector in balance equations for density, momentum,
and energy. Retaining the first order term ($\varepsilon$) in the
latter expansions, we come to the Navier-Stokes equations, while
next-order corrections are known as the Burnett ($\varepsilon^2 $)
and the super-Burnett ($\varepsilon^3 $) corrections \cite{chapman}.

However, as it was first demonstrated by Bobylev for Maxwell's
molecules \cite{bobylev1}, even in the simplest case
(one-dimensional linear deviation from global equilibrium) the
Burnett and the super-Burnett hydrodynamics violate the basic
physics behind the Boltzmann equation. Namely, sufficiently short
acoustic waves are increasing with time instead of decaying.
 This instability contradicts the
$H$-theorem, since all near-equilibrium perturbations must decay.
This creates difficulties for an extension of hydrodynamics, as
derived from a microscopic description, into a highly
non-equilibrium domain where the Navier-Stokes approximation is
inapplicable.

Recently, Bobylev suggested a different viewpoint on the problem of
Burnett's hydrodynamics \cite{Bo2006}. Namely, violation of
hyperbolicity can be seen as a source of instability. We remind that
Boltzmann's and Grad's equations are hyperbolic and stable due to
corresponding $H$-theorems. However, the Burnett hydrodynamics is
not hyperbolic which leads to no $H$-theorem. Bobylev \cite{Bo2006}
suggested to stipulate hyperbolization of Burnett's equations which
can also be considered as a change of variables. In this way
hyperbolically regularized Burnett's equations admit the $H$-theorem
(in the linear case, at least) and stability is restored.

Inspired by this study, in our recent paper \cite{cokk} (referred as
CKK hereafter), we have considered the simplest nontrivial example -
linearized Grad's moment equation in one spatial dimension - and
demonstrated that, upon a certain transformation, the exact (to all
orders in Knudsen number) hydrodynamic equations are manifestly
hyperbolic and stable. Thus, the first complete answer to what is
the structure of the extended hydrodynamics was obtained.

In this paper, we extend the CKK result to three-dimensional
linearized Grad's equations. In addition we prove the existence of
an $H$-function. The paper is organized as follows: In
Sec.~\ref{sec:1}, through a Dynamic Invariance Principle
\cite{GK92,GK94,Ka2005}, we derive equations of linear exact
hydrodynamics. In Sec.\ \ref{sec:2} we demonstrate that exact
hydrodynamic equations are manifestly hyperbolic and dissipative.
Then, In Sec.~\ref{sec:3} we stress explicitly how the stability
of hydrodynamic equations, and therefore the existence of an
$H$-theorem, arises as an interplay between these two basic
ingredients of resulting hydrodynamics: dissipativity and
hyperbolicity. Finally, a conclusion is given in Sec.\
\ref{sec:4}.

\section{Hydrodynamics from the linearized Grad system}
\label{sec:1}

\subsection{Linearized Grad's equations in $k$-space}
The thirteen moments linear Grad system consists of 13 linearized
PDE's giving the time evolution of the hydrodynamic fields (density
$\rho$, velocity vector field $\mbf{u}$, Temperature $T$) and of
higher order distinguished moments: five components of the symmetric
traceless stress tensor $\bfsigma$ and three components of the heat
flux $\q$ \cite{Grad}.

Point of departure is the Fourier transform of the linearized
three-dimensional Grad's thirteen-moment system:
 \beas{1b}
 \partial_t \rho_k &=& -i\k\cdot\u_k, \\
 \partial_t \u_k &=& -i\k \rho_k - i \k T_k - i \k\cdot\bfsigma_k,  \\
 \partial_t T_k &=& -\frac{2}{3}i \k\cdot(\u_k + \q_k), \\
 \partial_t \bfsigma_k &=& -2 i \stl{\k\u_k} -
 \frac{4}{5} i\stl{\k \q_k} -
 \bfsigma_k , \\
 \partial_t \q_k &=& -\frac{5}{2} i \k T_k - i \k\cdot\bfsigma_k -
 \frac{2}{3}\q,
\eeas where $\k$ is the wave vector, $\rho_k$, $\mbf{u}_k$ and $T_k$
are the Fourier components for density, average velocity and
temperature characterizing deviations from the equilibrium state,
respectively, and $\bfsigma_k$ and $\q_k$ are the nonequilibrium
traceless symmetric stress tensor ($\stl{\bfsigma}=\bfsigma$) and
heat flux vector components, respectively. The overline bar denotes
the traceless symmetric part of a 2nd rank tensor $\mbf{a}$,
$\stl{\mbf{a}}\equiv \frac{1}{2}(\mbf{a}+\mbf{a}^T)-\frac{1}{3}{\rm
tr}(\mbf{a}){\bf I}$ with unity matrix ${\bf I}$. The system
(\ref{1b}) provides the time evolution equations for a set of
hydrodynamic (locally conserved) fields $[\rho,\mbf{u},T]$ coupled
to the nonhydrodynamic fields $\bfsigma$ and $\q$. The goal is to
reduce the number of equations in (\ref{1b}) and to arrive at a
closed system for the hydrodynamic fields only.

To this end, it is common practice to decompose the vectors and
tensors into parallel (longitudinal) and orthogonal (lateral)
parts with respect to the wave vector, because the fields are
rotationally symmetric around any chosen direction $\k$. We
introduce a unit vector in the direction of the wave vector,
$\e=\k/k$, $k=|\k|$, and the corresponding decomposition, $\u_k =
u_k^\| \,\e + \u_k^\perp$, $\q_k = q_k^\| \,\e + \q_k^\perp$, and
$\bfsigma_k = \frac{3}{2} \sigma_k^\| \stl{\e\e} +
2\bfsigma_k^\perp$, where $\e\cdot\u_k^\perp=0$,
$\e\cdot\q_k^\perp=0$, and $\e\e:\bfsigma_k^\perp=0$.

Upon inserting the above decomposition into (\ref{1b}), and using
identities, $\stl{\e\e}\cdot\e=(2/3)\e$,
$\e\e:\stl{\e\e}=\stl{\e\e}:\stl{\e\e}=2/3$, we obtain the
following two closed sets of equations for the longitudinal and
lateral modes,

\bea
 \partial_t \rho_k &=& -i k \,u_k^\|,\nonumber \\
 \partial_t u_k^\| &=& -i k \rho_k - i k T_k - i k \sigma_k^\|, \nonumber\\
 \partial_t T_k &=& -\frac{2}{3}i k (u_k^\| + q_k^\|),\nonumber\\
 \partial_t \sigma_k^\| &=& -\frac{4}{3} i k u_k^{\|} - \frac{8}{15} i k q_k^{\|}
 - \sigma_k^\|, \nonumber \\
 \partial_t q_k^\| &=& -\frac{5}{2} i k T_k - i k \sigma_k^\|  - \frac{2}{3} q_k^\|,
 \label{reducedsetA}
\eea and
\bea
 \partial_t \u_k^\perp &=& - i k\, \e\cdot\bfsigma_k^\perp,\nonumber \\
 \partial_t \bfsigma_k^\perp &=& -i k \stl{\e\u_k^\perp} - \frac{2}{5} i k \stl{\e\q_k^\perp} -
  \bfsigma_k^\perp,\nonumber \\
 \partial_t \q_k^\perp &=& - i k \,\e\cdot\bfsigma_k^\perp - \frac{2}{3} \q_k^\perp.
 \label{reducedsetB}
\eea

Equations (\ref{reducedsetA}) and (\ref{reducedsetB}) are a
convenient starting point to derive closed equations for the
hydrodynamic fields. To this end, the Chapman-Enskog method amounts
to eliminating the time derivatives of the stress tensor and of the
heat flux in favor of spatial derivatives of the hydrodynamic fields
of progressively higher order. It had already been noted earlier
\cite{Ka2005} that we can express the stress tensor and the heat
flux vector  linearly in terms of the locally conserved fields by
introducing six, yet unknown, scalar functions $A(k),\dots,Z(k)$ for
the longitudinal part:
 \beas{introAZ}
 \sigma_k^\| &=& i k A u_k^\| - k^2 B \rho_k -k^2 C T_k, \\
  q_k^{\|} &=& i k X \rho_k + i k Y T_k - k^2 Z u_k^\|,
 \eeas
 and, respectively, two functions $D(k)$ and $U(k)$ for the
 transversal component,
 \beas{introDU}
 \bfsigma_k^\perp &=& i k D \stl{\e\u_k^\perp}, \\ 
 \q_k^{\bot} &=& -k^2 U \u_k^\perp, 
\eeas where the expressions for the longitudinal components share
their form with the one-dimensional CKK case. Note that the
functions introduced should be regarded as exact summation of the
Chapman-Enskog expansion which amounts to expanding these functions
into powers of $k^2$ and deriving coefficients of this expansions
from a recurrent (nonlinear) system, cf. CKK and \cite{Ka2005}). We
do not dwell on this here since we shall use a more direct way to
evaluate functions $A,\dots,Z,D,U$ in the sequel.

Finally, using expressions (\ref{introAZ}) and (\ref{introDU}) in
(\ref{reducedsetA}), (\ref{reducedsetB}) and denoting as
$x_k=(\rho_k,u_k^\|,T_k,\mbf{u}_k^\perp)$ the vector of the
hydrodynamical variables, the equations of hydrodynamics can be
written in a compact form using a block-diagonal matrix $\Mk$,

\be
 \partial_t x_k = \Mk \,x_k, \qquad \Mk =
\left(\begin{array}{cc}
  \Mk^{\|} & 0 \\
  0& \Mk^\perp \\
  \end{array}\right),
  \label{M3D}
\ee with
 \be
 \Mk^\| = \left(\begin{array}{ccc}
  0 & -i k & 0 \\
  - i k( 1\! -\! k^2 B) &  k^2 A & -i k ( 1\! -\! k^2 C) \\
  \frac{2}{3}k^2 X & - \frac{2}{3} i k ( 1\! -\! k^2 Z) &\frac{2}{3}k^2Y
 \end{array}\right),
 \label{MAparall}
\ee
 and
 \be
 \Mk^\perp =k^{2}D \left(%
\begin{array}{cc}
  1 & 0 \\
  0 & 1 \\
\end{array}%
\right),
 \label{MAperp}
\ee where the unit matrix is written in an (arbitrarily) fixed basis
in the two-dimensional subspace of vectors $\u_k^\perp$.
 As follows from an immediate comparison with
CKK, and due to the apparently useful notation, the matrix $\Mk^\|$
providing the evolution of the longitudinal modes, is exactly
identical with the corresponding matrix (denoted as ${\mbf M}$ in
CKK) for the one-dimensional case, where lateral modes are absent.
The twice degenerated transversal (shear) mode is decoupled from the
longitudinal modes. As a direct consequence, also the invariance
equations to be discussed next, which will provide us with a set of
nonlinear algebraic equations for the unknown functions $A$--$Z$,
divide into two sub-blocks which can be solved separately.

\subsection{Invariance Equations}

In order to evaluate functions $A,\dots,Z,D,U$, we make use of the
dynamic invariance principle (DIP) \cite{GK92,GK94,Ka2005}. Making
use of DIP in just the same way as for the one-dimensional case
(CKK) leads to two independent sets of invariance equations for
the functions $A(k)$--$Z(k)$. We find that the first set (six
coupled quadratic equations for $A,B,C$ and $X,Y,Z$) is identical
to the one already presented, cf. CKK, Eq.\ (17).

For the transversal modes, the invariance condition reads,
\begin{eqnarray}
\frac{\partial  \bfsigma_k^\perp}{\partial
\u_k^\perp}\cdot(-ik\e\cdot \bfsigma_k^\perp)=\partial_t
\bfsigma_k^\perp,\nonumber\\
\frac{\partial  \q_k^\perp}{\partial \u_k^\perp}\cdot(-ik\e\cdot
\bfsigma_k^\perp)=\partial_t \q_k^\perp, \label{invDU}
\end{eqnarray}
where the time derivative in the left hand side is evaluated by
chain rule using $\partial_t\u_k^\perp$. Substituting the functions
(\ref{introDU}) into (\ref{invDU}), and requiring that the
invariance condition is valid for any $\u_k^\perp$, we derive two
coupled quadratic equations for the functions $D$ and $U$ which can
be cast into the following form:
\begin{eqnarray}
  && 15 k^4 D^3 + 25 k^2 D^2 + (10+21 k^2) D + 10 = 0, \nonumber \\
&&U=-\frac{3D}{2+3k^2D}. \label{DU}
\end{eqnarray}

\begin{figure}[t]
\includegraphics[width=\width]{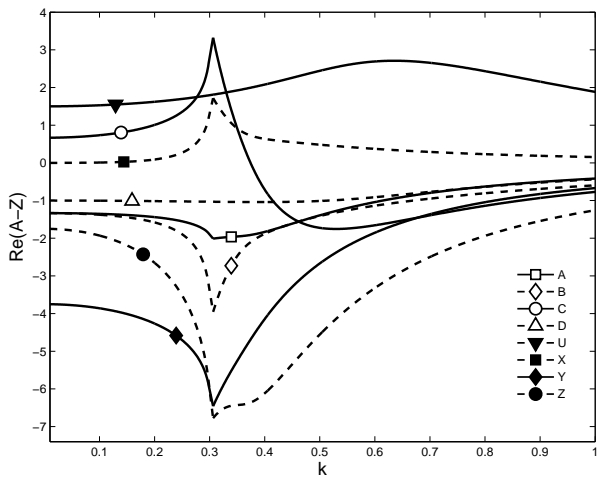}
\caption{Real parts of coefficients $A$ to $Z$ solving the
invariance equations, CKK (Eq.~17) supplemented with (\ref{DU}).
}\label{Re_AZ_3D}
\end{figure}

Solution of the cubic equation (\ref{DU}) with the initial condition
$D(0)=-1$ matches the Navier-Stokes asymptotics and was found
analytically for all $k$. This solution is real-valued and is in the
range
 $D(k)\in
[-1.04,0]$, whereas $U(k)\in[0,2.72]$.
 The functions corresponding to the longitudinal part of the
 system
have been obtained numerically in CKK. Because $D$ and $U$ are
real-valued, we show in Fig.~\ref{Re_AZ_3D} the real parts for all
coefficients, while their nonvanishing imaginary parts still
coincide with those shown in CKK Fig.~4.

\begin{figure}[t]
\includegraphics[width=\width]{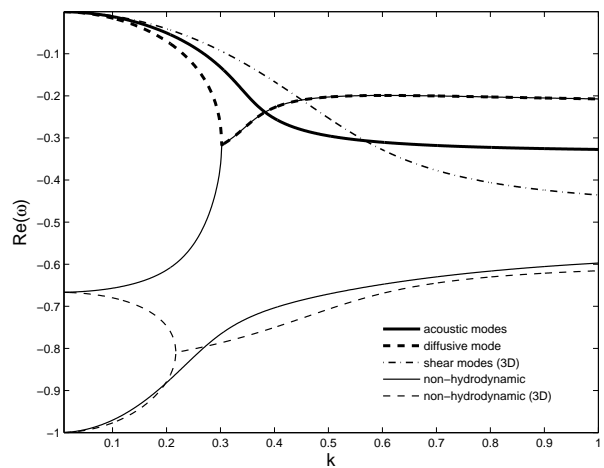}
\caption{Dispersion relations $\omega(k)$ for the linearized
Grad's system using projected variables, Eqs.\ (\ref{reducedsetA})
and (\ref{reducedsetB}). The five hydrodynamic modes (diffusive,
twice degenerated shear, and two complex-conjugated acoustic
modes), as well as the eight non-hydrodynamic modes are presented
as a function of $k$. While the acoustic mode is complex-valued
for all $k$, the remaining modes become complex-valued beyond the
two visible bifurcation points (at $k'_{\rm c}\approx 0.2175$ and
$k_c\approx 0.303$) For $k<k_c'$ the non-hydrodynamic (3D) modes
are degenerated two and four times, respectively, corresponding to
the two and four components of $\q_k^\perp$ and
$\bfsigma_k^\perp$. }\label{mkfigover3D}
\end{figure}

The dispersion relations $\omega(k)$ for the five hydrodynamic modes
are then calculated by inserting these coefficients into the roots
of characteristic equation ${\rm det}\left( \Mk - \omega{\cal I}
\right)=0$, where ${\cal I}$ is a $5\times 5$ unit matrix.
Analogously, the dispersion relations for the remaining
non-hydrodynamic modes follow from eight (remaining) eigenvalues of
(\ref{reducedsetA}), (\ref{reducedsetB}) with (\ref{introAZ}),
(\ref{introDU}). All 13 modes are presented in Fig.\
\ref{mkfigover3D}. The resulting hydrodynamic spectrum consist of
five modes: the acoustic mode, $\omega_{\rm ac}(k)$, represented by
two complex-conjugated roots, the real-valued thermal (diffusive)
mode, (both modes already occurring in the one-dimensional case) and
a twice-degenerated real-valued
shear mode 
(cf.\ Fig.\ \ref{mkfigover3D}). The occurrence of a real-valued
shear mode confirms a more general result: in the linear regime, the
shear mode never undergoes damped oscillations. Same as in the
one-dimensional case,  a critical point  in the hydrodynamic
spectrum occurs at $k_c\approx0.303$, where the thermal mode
intersects a non-hydrodynamical branch of the original Grad system.
Hence, same conclusions hold here: for $k\ge k_{c}$, the CE method
does not recognize any longer the resulting diffusive branch as an
extension of a hydrodynamic branch. Figure~\ref{mkfigover3D} further
shows the eight (all degenerated) non-hydrodynamic modes, which in
opposite to the one-dimensional case (offering two non-hydrodynamic
modes) also exhibit a critical $k$ at $k_c'\approx 0.2175$.

To summarize, exact hydrodynamics as derived from invariance
condition (or, equivalently, by the complete summation of the CE
expansion as demonstrated in CKK (cf. also \cite{Ka2005}) extends
up to a finite critical value $k_{c}$, in full agreement with the
one-dimensional case. No stability violation occurs, unlike in the
finite-order truncations thereof. Next, we address the question
about hyperbolicity of exact hydrodynamics in the present
three-dimensional case.

\section{Hyperbolicity of exact hydrodynamics}
\label{sec:2}

Distinguishing between the real ($\Ree_{k}$) and imaginary
($\Imm_{k}$) parts of matrix $\Mk$ (\ref{M3D}), we can write the
equation of hydrodynamics  conveniently as \bea\label{short}
 \partial_{t} x_k  &=& [\MR_{k}-i\MI_{k}]\,x_k,
\eea
\begin{equation}
\MR_{k}=\left(\begin{array}{cc}
  {\rm Re}(\Mk^{\|}) & 0 \\
  0& \Mk^\perp \\
  \end{array}\right),\ -\MI_{k}=\left(\begin{array}{cc}
  {\rm Im}(\Mk^{\|}) & 0 \\
  0& 0\\
  \end{array}\right).
\end{equation}

The system (\ref{short}) is hyperbolic and stable if we can find a
transformation of the hydrodynamic fields, ${x}_k' = \Tk x_k$, where
$\Tk$ is a real-valued matrix, such that, for the transformed
matrices $\Mk'=\Tk\Mk\Tk^{-1}$ it holds
\begin{enumerate}
\item[(i)] $\MR'_{k}={\rm Re}(\Mk')$ and $\MI'_{k}={\rm Im}(\Mk')$ are
symmetric, and \item[(ii)] $\MR'_{k}$ has non-positive eigenvalues.
\end{enumerate}

Due to the block-diagonal structure of (\ref{M3D}) as well as to
the fact that CKK has solved the problem of finding a
transformation with the desired properties for the one-dimensional
case, the transformation exists also in the three-dimensional
case, and has the following form:
 \be
 \Tk = \left(\begin{array}{cc}
  \Tk^{\|} & 0 \\
  0& \Tk^\perp \\
  \end{array}\right),
 \label{T3D}
\ee
 where $\Tk^{\|}$ is explicitly given by CKK Eqs.~(25)--(27) in terms of $k$, $A$--$C$ and $X$--$Z$, and
 \begin{equation}
 \Tk^\perp=\left(%
\begin{array}{cc}
  1 & 0 \\
  0 & 1 \\
\end{array}%
\right).
 \end{equation}

Thus, the transformation $\Tk$ (\ref{T3D}) symmetrizes $\Mk$ and
renders the exact hydrodynamic equations manifestly hyperbolic.
Furthermore, the transform $\Tk$ contains only even powers of $k$,
 because the same is true for the coefficients $A$--$Z$.

 The five eigenvalues $\lambda_{1-5}$ of $\Ree'_{k}$ (or,
equally, of $\Ree_{k}$), are
\begin{equation}
\lambda_1=0, \;\;\;\lambda_2=k^2 A, \;\;\; \lambda_3=\frac{2}{3}k^2
Y, \;\;\;\lambda_{4,5}=k^2D. \label{18}
\end{equation}

From the analysis of the previous section, where we solved for
coefficients $A$, $D$, and $Y$ appearing in (\ref{18}), cf.
Fig.~\ref{Re_AZ_3D}, it follows that all the eigenvalues
$\lambda_{1-5}$ are non-positive for all $k$. Note that the matrix
$\MR'_{k}$ is diagonal with the diagonal elements (\ref{18}).

\section{H-theorem for exact hydrodynamics}
\label{sec:3}

Finally, the hyperbolic structure straightforwardly implies an
$H$-theorem for the exact hydrodynamics (the same holds for any
lower order approximation, if they are obtained according to the
method presented in CKK). Note that, due to linearity of the
system (\ref{1b}), the choice of a proper $H$-functional is not
unique. We follow Bobylev \cite{Bo2006}, and consider an
$H$-function -- in terms of the transformed hydrodynamic fields --
defined as:
\begin{equation}
H=\frac{1}{2}  \int
\left[\rho'^{2}(\r,t)+u'^{2}(\r,t)+T'^{2}(\r,t)\right] d^3 r.
\end{equation}
Here, hydrodynamic fields $x'(\r,t)$ are defined through inverse
Fourier transform of the fields $x'_k$. Note that $x'(\r,t)$ are
real-valued because the real-valued transformation $\Tk$ is an even
function of $k$, $\Tk=\Tmk$. Therefore,
\begin{equation}
H=\frac{1}{2}  \int
\left[\rho'_k\rho'_{-k}+\u'_k\cdot\u'_{-k}+T'_kT'_{-k}\right] d^3 k,
\end{equation}
which we abbreviate as $H=\frac{1}{2}\ave{{x}'_k,{x}'_{-k}}$.
Thus, \bea
\partial_{t}H&=& \frac{1}{2}(\ave{x'_k,\partial_tx'_{-k}} + \ave{\partial_tx'_k,x'_{-k}})\nonumber\\
&=& -\frac{1}{2}i(\ave{x'_k,\MI'_{-k}x'_{-k}} +
\ave{x'_{-k},\MI'_{k}x'_{k}})\nonumber\\
 && + \frac{1}{2}(\ave{x'_k,\MR'_{-k}x'_{-k}} +
\ave{x'_{-k},\MR'_{k}x'_{k}}). \eea

Since $\MI'_k$ is an odd function of $k$, $\MI'_{-k}=-\MI'_k$, terms
containing $\MI'$ cancel out, and we have, owing to the fact that
$\MR'$ is even function of $k$ ($\MR'_{-k}=\MR'_k$),
\begin{equation}
\partial_t H=\sum_{s=1}^5 \int\lambda_s|x'_{s,k}|^2d^3k\le0.
\end{equation}
Thus, we have proved the $H$-theorem for the exact hydrodynamics for
$k<k_c$ (at $k=k_c$, the eigenvalues $\lambda_2$ and $\lambda_3$
become complex-valued, as discussed above).

\section{Conclusions}
\label{sec:4}

In this paper, we have considered derivation of exact
hydrodynamics from linearized three-dimensional Grad's system. The
main finding is that the exact hydrodynamic equations (summation
of the Chapman-Enskog expansion to all orders) are manifestly
hyperbolic and stable, thereby extending the previous CKK result
\cite{cokk}. To the best of our knowledge, this is the first
complete answer of the kind. The study supports the recent
suggestion of Bobylev on the hyperbolic regularization of
Burnett's approximation. We have also demonstrated, by a direct
computation, the $H$-theorem for the quadratic entropy function.

\subsection*{Acknowledgment}

I.V.K. gratefully acknowledges support by BFE Project 100862  and by
CCEM-CH. M.K. acknowledges support through grants
NMP3-CT-2005-016375 and FP6-2004-NMP-TI-4 STRP 033339 of the
European Community.

\end{document}